\documentclass[numberedappendix,iop]{emulateapj}

\usepackage{dcolumn}
\usepackage{bm}
\usepackage{epsf}
\usepackage{verbatim}
\usepackage{amsmath}
\usepackage{amssymb}
\usepackage{tikz}
\tikzset{fontscale/.style = {font=\relsize{#1}}}
\usepackage{color}
\usepackage{tipa}
\usepackage{verbatim}
\usepackage{amsfonts}
\usepackage{xcolor}
\definecolor{darkgreen}{rgb}{0.0,0.5,0.0}

\usepackage[colorlinks,citecolor=darkgreen]{hyperref}

\usepackage[makeroom]{cancel}

\DeclareSymbolFont{matha}{OML}{txmi}{m}{it}
\DeclareMathSymbol{\varv}{\mathord}{matha}{118}

\newcommand{\ie}{\emph{i.e.} }
\newcommand{\eg}{\emph{e.g.,} }

\newcommand{\be}{\begin{equation}}
\newcommand{\ee}{\end{equation}}
\newcommand{\bea}{\begin{equation*}}
\newcommand{\eea}{\end{equation*}}
\newcommand{\beqr}{\begin{eqnarray} \nonumber}
\newcommand{\eeqr}{\end{eqnarray}}
\newcommand{\beqrb}{\begin{eqnarray}}
\newcommand{\eeqrb}{\nonumber \end{eqnarray}}
\newcommand{\fin}{\mbox{ .}}
\newcommand{\comma}{\mbox{ ,}}

\newcommand{\const}{\mbox{const}}

\newcommand{\gama}{$\gamma$}

\newcommand{\nab}{\nabla}

\newcommand{\pr}{\partial}

\newcommand{\Dim}{{N}}
\newcommand{\Nfs}{{M}}
\newcommand{\myQ}{{q}}
\newcommand{\myq}{{\Lambda}}
\newcommand{\DoF}{{\nu}}
\newcommand{\DoFInverse}{{\nu^{-1}}}
\newcommand{\Se}{{s_E}}
\newcommand{\SP}{{s_p}}
\newcommand{\SpIso}{s_p^{\text{(iso)}}}
\newcommand{\SeIso}{s_E^{\text{(iso)}}}
\newcommand{\Ad}{{\Gamma_{\text{ad}}}}
\newcommand{\np}{\mathcal{N}}
\newcommand{\db}{ \beta_r}

\newcommand{\myComp}{\mathcal{R}}
\newcommand{\myD}{\mathfrak{D}}

\newcommand{\appropto}{\mathrel{\vcenter{
  \offinterlineskip\halign{\hfil$##$\cr
    \propto\cr\noalign{\kern2pt}\sim\cr\noalign{\kern-2pt}}}}}

\DeclareMathOperator\arctanh{arctanh}

\DeclareMathOperator*{\myK}{K}

\defcitealias{Keshet_2005}{KW05}
\newcommand{\KW}{{\citetalias{Keshet_2005}}}

\newcommand{\Arad}{A20}

\begin{document}

\title{Diffusive shock acceleration in $N$ dimensions}

\shorttitle{$N$-dimensional DSA}

\author{Assaf Lavi}

\author{Ofir Arad}

\author{Yotam Nagar}

\author{Uri Keshet}

\email{assafla@post.bgu.ac.il}
\email{ukeshet@bgu.ac.il}

\affil{Physics Department, Ben-Gurion University of the Negev, POB 653, Be'er-Sheva 84105, Israel}

\shortauthors{Lavi et al.}

\date{\today}

\begin{abstract}
Collisionless shocks are often studied in two spatial dimensions (2D), to gain insights into the 3D case.
We analyze diffusive shock acceleration for an arbitrary number $N\in\mathbb{N}$ of dimensions.
For a non-relativistic shock of compression ratio $\mathcal{R}$, the spectral index of the accelerated particles is $s_E=1+N/(\mathcal{R}-1)$; this curiously yields, for any $N$, the familiar $s_E=2$ (\ie equal energy per logarithmic particle energy bin) for a strong shock in a mono-atomic gas.
A precise relation between $s_E$ and the anisotropy along an arbitrary relativistic shock is derived, and is used to obtain an analytic expression for $s_E$ in the case of isotropic angular diffusion, affirming an analogous result in 3D.
In particular, this approach yields $s_E = (1+\sqrt{13})/2 \simeq 2.30$ in the ultra-relativistic shock limit for $N=2$, and $s_E(N\to\infty)=2$ for any strong shock.
The angular eigenfunctions of the isotropic-diffusion transport equation reduce in 2D to elliptic cosine functions,
providing a rigorous solution to the problem; the first function upstream already yields a remarkably accurate approximation.
We show how these and additional results can be used to promote the study of shocks in 3D.
\end{abstract}


\maketitle

\section{introduction}
\label{sec:Intro}

Collisionless shocks are known to accelerate charged particles to ultra-relativistic energies in a wide range of astronomical systems.
Diffusive shock acceleration (DSA) is a first-order Fermi mechanism \citep{Fermi49, Bell_1978}, thought to be responsible for this process.
DSA can explain, under certain assumptions, the power-law spectra of high-energy particles inferred in various astrophysical phenomena; for reviews, see \citet{drury1983introduction, Blandford_Eichler_87, sironi2015relativistic}.
While most DSA studies focus on $N=3$ spatial dimensions \citep[3D; but see the 1D analysis of][discussed below]{keshet2017analytic}, valuable insights and further analytic progress is possible in other dimensions, in particular low $N$ where the problem becomes simpler and more easily tractable computationally by ab-initio simulations.

DSA involves energetic particles, scattered by electromagnetic modes, bouncing between the upstream and downstream sides of a shock, gradually gaining energy in each cycle.
The process is still not understood from first principles. A self-consistent model would need to simultaneously account for the injection and acceleration of the particles, their scattering off electromagnetic irregularities, and the formation of these irregularities by the bulk flow and by the accelerated particles themselves.
One approach to the problem is the test-particle approximation, evolving the particle distribution function (PDF) $f$ by adopting some ansatz for the scattering mechanism and neglecting the backreaction of the accelerated particles on the shock and on the scattering medium.

This approach was used to derive the energy spectral index,
\begin{equation}
s_E \equiv -\frac{d\ln n(E)}{d\ln E} = 1-\Dim-\frac{d\ln f}{d\ln E} \comma
\end{equation}
where $n(E)$ is the specific particle density. In non-relativistic shocks in 3D \citep{Krymskii_1977,1977ICRC...11..132A,Bell_1978,1978ApJ...221L..29B},
\begin{equation}  \label{eq:ClassicSE3D}
\Se \simeq \frac{\myComp+2}{\myComp-1}
\end{equation}
depends only on the shock compression ratio $\myComp$, although this result does not necessarily apply when scattering is highly anisotropic \citep{KeshetEtAl19}.

For sufficiently isotropic scattering around a strong shock in an ideal mono-atomic gas, $\myComp\to4$ then implies that $\Se\simeq 2$.
While this approach does not address possible non-linear effects on the shock and the scattering modes
\citep[see][for reviews]{Blandford_Eichler_87, malkov2001nonlinear,2016MNRAS.456.3090E},
it is consistent with a wide range of observations.
The flat, $\Se\to 2$ spectrum is peculiar in its logarithmic energy divergence, and one may ask if it is a coincidence that this spectrum appears to be most relevant in nature.
It is interesting to generalize the result to dimensions other than three, and to ask for example whether the emerging flat spectrum is general or unique to a mono-atomic gas in 3D.

Relativistic shocks raise additional questions, which can benefit from an analysis with a different number of dimensions.
DSA is more complicated in a relativistic shock, as the PDF can no longer be approximated as isotropic. Assuming small-angle scattering, parameterized by an angular diffusion function $\myD$,
analyses of DSA in relativistic shocks by numerical \citep[\eg][]{bednarz1998energy,achterberg2001particle}, semi-analytic \citep[][]{kirk1987acceleration, heavens1988relativistic, Kirk_2000, Keshet06}, and analytic \citep[][henceforth \KW]{Keshet_2005} methods  found a spectral index  $\Se\simeq 22/9\simeq 2.22$ in the ultra-relativistic shock limit for isotropic diffusion.

This result broadly agrees with observations of systems associated with non-magnetized relativistic shocks, such as \gama-ray burst (GRB) afterglows \citep[\eg][and references therein]{Waxman_06} and possibly also jets in BL-Lac objects.
An analysis of $\sim 300$ GRB afterglows found $\Se\simeq 2.25$ as most likely, but with a broad, $\Se=2.36\pm0.59$ Gaussian distribution \citep{curran2010electron}, probably due a long tail of soft-spectrum GRBs \citep{Ryan_2015}. Focusing only on short GRB afterglows, a distribution of $\Se=2.43^{+0.36}_{-0.28}$ was found from a sample of 38 such events \citep{fongbergerGRB}.
Similarly, jets in BL-Lac objects, in which the polarization pattern is consistent with shock-generated magnetic fields, show $\Se=2.28\pm 0.06$ \citep{hovatta2014mojave}.
It is thought that magnetized relativistic shocks cannot efficiently accelerate particles \citep[\eg][]{Kirk1989, Begelman1990, BallardHeavens91, OstrowskiBednarz02, sironi2009particle}
although in the extreme limit of pulsar wind nebulae (PWN), the highly magnetized termination shock is thought to accelerate an extremely hard, $\Se\simeq 1$ spectrum, possibly through diffusive shock acceleration \citep[\eg][and Arad et al., in prep., henceforth \Arad]{fleishman2007diffusive}.
Insights obtained from DSA in other dimensions may shed light on these phenomena.

The problem of DSA in relativistic shocks was not rigorously solved, and the dependence of $\Se$ on the diffusion function is not yet entirely clear.
For instance, a precise relation exists between the spectrum and the particle anisotropy at the shock front; for isotropic diffusion in 3D, this leads to (\KW)
\begin{equation} \label{eq:S3D}
\Se \simeq \frac{\beta_u+2\beta_d-2\beta_u\beta_d^2+\beta_d^3}{\beta_u-\beta_d} \comma
\end{equation}
where $\beta$ is the fluid velocity with respect to the shock, normalized to the speed $c$ of light, and upstream (downstream) quantities are labelled with subscript $u$ (subscript $d$), written henceforth only when necessary.
This result is quite sensitive to the angular form of the diffusion function \citep[][and A20]{Keshet06}, although, interestingly, not to local feedback from the relativistic particles \citep{NagarKeshet19}.
It would be useful to generalize Eq.~(\ref{eq:S3D}) , in particular to 2D, where ab-initio simulations are increasingly capable of resolving particle acceleration.

In this study, we generalize the test-particle analysis of DSA to $N\neq 3$ dimensions.
As the preceding discussion indicates, analyzing such DSA is useful for several reasons.
First, low-dimensional studies are often essential due to the complexity of the 3D case.
Indeed, as resolving a 3D collisionless shock is at present prohibitively expensive computationally, much of the progress in the field has relied on the analysis of 1D or 2D systems.
The $N=2$ case is especially important, as 2D shocks manifest key processes relevant to 3D, yet can be substantially evolved numerically \citep[\eg][]{2008ApJ...673L..39S, martins2009ion, Keshet09, sironi2009particle, sironi2013maximum, caprioli2014simulations, caprioli2017chemical}.
Indeed, 3D experiments are often interpreted based on 2D simulations \citep[\eg][]{takabe2008high, liu2011collisionless, kuramitsu2011time, haberberger2012collisionless}.
Second, $N\neq3$ studies provide valuable insights which are inherently inaccessible in a 3D framework.
For example, DSA in 1D is uniquely independent of the scattering function, so its curious behavior in the ultra-relativistic limit may be indicative of 3D shocks, in which the scattering function is important but poorly constrained \citep{keshet2017analytic}.
As another example, we show that the flat spectrum ($\Se=2$, for an ideal, mono-atomic gas) in the non-relativistic, strong shock limit is independent of $N$, suggesting that the corresponding logarithmic energy convergence is not coincidental.
Third, low-dimensional analyses may be effectively applicable to some physical systems. For instance, in a strongly magnetized parallel shock, magnetic confinement can render the system effectively 1D.
Fourth, experimental work has recently managed to effectively realize 2D shocks, in systems such as gas tubes \citep[\eg][]{skews2015experimental} and shallow-water analogues \citep[\eg][]{foglizzo2012shallow}.
Finally, our $N\neq3$ results can be used for code development and verification, and for pedagogical purposes.

The paper is organized as follows.
In \S\ref{sec:nDim}, we outline the DSA problem in $\Dim$ dimensions.
The problem is solved for non-relativistic shocks in \S\ref{sec:non-rel}.
We specify to $\Dim=2$ in \S\ref{sec:2D}, calculating the spectrum and PDF for an arbitrarily relativistic shock in several analytic and semi-analytic methods, with an emphasis on the ultra-relativistic shock limit.
The analysis of relativistic shocks is generalized to $\Dim$ dimensions in \S\ref{sec:RelND}.
Our results are summarized and discussed in \S\ref{sec:Summary}.
In appendix \S\ref{sec:DiffusionEq}, we derive the transport equation in 2D.
In \S\ref{sec:MJ}, we derive the Maxwell--J\"uttner distribution for an arbitrarily-relativistic gas in 2D, subsequently used in \S\ref{sec:JS_eos} to derive the 2D J\"uttner--Synge (JS) equation of state (EOS) and the corresponding shock jump conditions.
Appendix \S\ref{sec:err} details the convergence properties of our results and our error estimation.

\section{DSA in $\Dim$ dimensions}\label{sec:nDim}

In this section, we present the DSA problem in a general setting with $\Dim$ spatial dimensions.
In \S\ref{sec:Jump}, we discuss the $\Dim$-dimensional shock jump conditions.
We lay out the $\Dim$-dimensional DSA setup in \S\ref{sec:formalism}.

\subsection{Shock jump conditions}\label{sec:Jump}

In a non-relativistic fluid, the adiabatic index of an ideal gas is given by $\Ad=1+2\DoFInverse$ \citep[\eg][]{ryden2016dynamics},
where $\DoF$ is the effective number of particle degrees of freedom. The Rankine-Hugoniot jump conditions \citep[\eg][]{landau1959fluid} hold in any $\Dim$, so the compression ratio of a strong non-relativistic shock is given by
\begin{equation}\label{eq_rNR}
\myComp\equiv\frac{\beta_u}{\beta_d} \to \frac{\Ad+1}{\Ad-1} = 1+\DoF\fin
\end{equation}
In an ultra-relativistic fluid, $\Ad=1+\DoFInverse$, so here, for a strong shock \citep[\eg][]{keshet2017analytic}
\begin{equation}\label{eq_rUR}
\myComp\equiv \frac{\beta_u}{\beta_d} \to \frac{\mathcal{\epsilon}_d+P_u}{\mathcal{\epsilon}_u+P_d} \simeq \frac{\mathcal{\epsilon}_d}{P_d} \simeq \frac{1}{\Ad-1}=\DoF\comma
\end{equation}
where $\mathcal{\epsilon}$ is the internal energy density and $P$ is the pressure.

In a relativistic fluid, the adiabatic index is typically assumed to vary smoothly between the above non-relativistic and ultra-relativistic limits, according to the
JS EOS \citep{ANDP:ANDP19113390503,synge1957relativistic}. The phase-space particle distribution in such a fluid, known as the Maxwell-J\"uttner distribution, can be derived by minimizing the free energy, and is used to infer the JS EOS. Here, we focus on the 2D case, deriving the 2D Maxwell-J\"uttner distribution in  Appendix \S\ref{sec:MJ}, and the respective JS EOS and shock jump conditions in \S\ref{sec:JS_eos}.

\begin{figure}[h]
\centerline{\epsfxsize=8.8cm \epsfbox{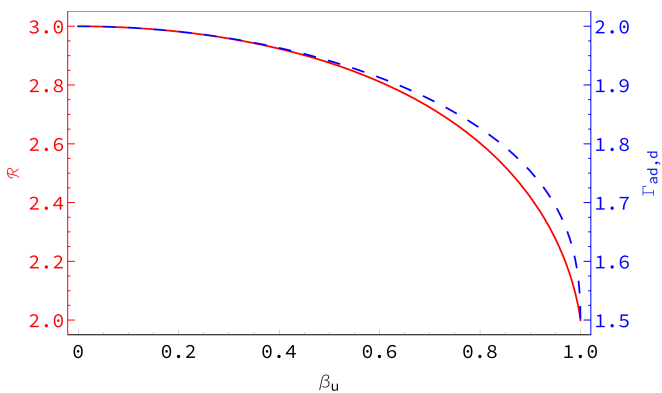}}
    \caption{
    The downstream adiabatic index $\Gamma_{\text{ad},d}$ (blue dashed curve with right axis)
    and the compression ratio $\myComp$ (red solid curve with left axis)
    are shown for a strong shock as a function of the normalized upstream velocity $\beta_u$, for the JS EOS in 2D.}
    \label{fig:AdComp}
\end{figure}

Summarizing the results of Appendix \S\ref{sec:JS_eos}, the 2D JS EOS can be written in terms of the dimensionless inverse temperature, $\zeta\equiv mc^2/(k_B T)$, in the form
\begin{equation}\label{eq_ad}
    \Ad=2-\frac{1}{2+\zeta}\simeq
    \begin{cases}
     2-\frac{1}{\zeta }
     +O\left(\zeta^{-2}\right)& \zeta\gg 1\, ; \\
     \frac{3}{2}+\frac{\zeta }{4}
     +O\left(\zeta^2 \right) & \zeta\ll 1 \comma
    \end{cases}
\end{equation}
where $m$ is the particle mass, $T$ is the plasma temperature, and $k_B$ is the Boltzmann constant.
The non-relativistic and ultra-relativistic limits in Eq.~(\ref{eq_ad}) agree with the respective limits in Eqs.~(\ref{eq_rNR}-\ref{eq_rUR}), where $\nu=2$ for a mono-atomic gas in 2D.
Then, in the case of a strong shock in 2D, the EOS along with the conservation of mass, momentum, and energy fluxes across the shock, yield the jump condition
\begin{equation}\label{eq_jump_cond}
    1+\frac{\zeta_d +2}{(\zeta_d+1)\zeta_d}=(1-\beta _d \beta _u)\gamma_d \gamma_u \comma
\end{equation}
where $\zeta_d$ is derived as a function of $\beta_u$ as the positive root of a seventh-order polynomial provided in Eq.~(\ref{eq_zeta}). Here and henceforth, $\gamma\equiv (1-\beta^2)^{-1/2}$ is the fluid Lorentz factor.
The resulting downstream adiabatic index and shock compression ratio $\myComp$ are presented, as a function of $\beta_u$, in Figure \ref{fig:AdComp}.

\subsection{DSA setup}
\label{sec:formalism}

Consider DSA in $\Dim\geq2$ dimensions.
We work, as in most analytic studies, in the test-particle approximation.
More precisely, as the scattering function in a relativistic shock \citep[and sometimes even in a non-relativistic shock; see][]{KeshetEtAl19} is not rigorously known,
we simply work with a prescribed scattering function.
We avoid the injection problem, assuming that particles are injected at the shock front with sufficiently high energies to easily cross the shock.

Let $z_s$ be the oriented distance from the shock, in the shock frame (we henceforth omit the subscript $s$ unless necessary), such that the flow is in the positive $z$ direction and the shock is at $z=0$.
We assume that by averaging over constant $z$ planes, parallel to shock front, the resulting, reduced PDF, $f(z,\bm{p})$,  is time-independent, where $\bm{p}$ is the particle momentum.

Further assuming that plane-averaging leaves no preferred direction in the system, we arrive at the reduced
Lorentz-invariant, steady-state PDF $f(z,p,\mu)$.
Here, $\arccos(\mu)$ is the angle between momentum and flow directions.
In this mixed-frame, three-dimensional phase space,
$z$ is measured in the shock frame,
whereas $p$ and $-1\leq\mu\leq1$ are measured in the fluid frame.
Note that unlike the polar angle $\theta\equiv \arccos \mu$ in $N>2$ dimensions, the azimuthal angle $\phi\equiv\arccos\mu$ for $N=2$ is periodic.

Under the above assumptions, the particle momentum $p$ is much larger than any momentum scale in the system.
The lack of a characteristic scale then implies a power-law spectrum, so the PDF may be written as $f=\myQ(z,\mu)p^{-\SP}$, reducing the problem to determining the constant $\SP$ and the function $\myQ(z,\mu)$.
The momentum spectral index $\SP$ is related to the energy spectral index $\Se$ by
\begin{equation}
\SP=\Se+\Dim-1\fin
\end{equation}

\begin{figure}[h]
	\centering
\begin{tikzpicture}
    \node[anchor=south west,inner sep=0] at (0,0)
    {\centerline{\epsfxsize=7.5cm \epsfbox{"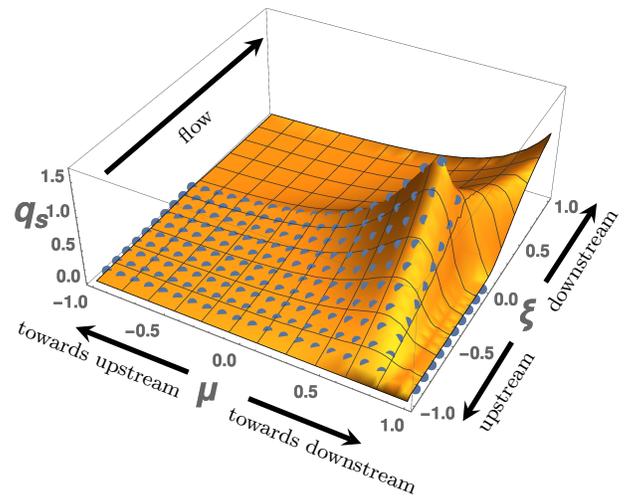"}}};
    \draw[line width=2pt,black,-stealth](7.2,1.25)--(6.55,0.2);
    \draw[line width=2pt,black,-stealth](7.65,2.05)--(8.33,3.1);
    \node[label={[label distance=0.5cm,text depth=-1ex,rotate=58]right:upstream}] at (6.32,-0.4) {};
    \node[label={[label distance=0.5cm,text depth=-1ex,rotate=58]right:downstream}] at (7.23,1.2) {};
    \draw[line width=2pt,black,-stealth](1.8,3.5)--(3.9,5.3);
    \node[label={[label distance=0.5cm,text depth=-1ex,rotate=42]right:flow}] at (2.2,3.6) {};
    \draw[line width=2pt,black,-stealth](2.9,0.85)--(1.4,1.45);
    \node[label={[label distance=0.5cm,text depth=-1ex,rotate=-22]right:towards upstream}] at (0,1.75) {};
    \draw[line width=2pt,black,-stealth](3.7,0.5)--(5.2,-0.1);
    \node[label={[label distance=0.5cm,text depth=-1ex,rotate=-22]right:towards downstream}] at (2.8,0.6) {};
\end{tikzpicture}
	\caption{
	Normalized shock-frame reduced PDF $q_s(\xi,\mu)$ of accelerated particles around a strong relativistic shock with covariant velocity $\gamma_u \beta_u=10$, with the JS EOS, in 2D.
Here, $\xi$ is defined by $z\propto \arctanh{\xi}$, such that negative (positive) $\xi$ correspond to the upstream (downstream) region, and $\xi=0$ is the shock front. The PDF is shown based on a numerical code \citep[orange surface;][]{NagarKeshet19} and on $M=10$ upstream eigenfunctions derived in \S\ref{sec:eigen2} (blue disks intersecting the surface).
The normalization (unit integral of $q_s$ over $\mu$ at $\xi=0$) is arbitrary.
	}
	\label{fig:distribution_map}
\end{figure}

Figure \ref{fig:distribution_map} illustrates the reduced PDF $q$ in its entire domain, in the shock frame.
In this example, we consider a strong, ultra-relativistic shock in 2D, with an isotropic angular diffusion function $D=\const.$ (defined formally in \S\ref{sec:formalism_2d}).
The shock-frame PDF, $q_s$, is plotted based both on a numerical finite-differences code \citep[][orange surface]{NagarKeshet19} and on our upstream eigenfunction expansion (see \S\ref{sec:eigen2}; blue discs intersecting the surface, shown in the upstream only).
The unbound coordinate $z$ is mapped onto a compact coordinate, $-1<\xi\equiv \tanh (D \gamma^3 z/c)<1$, so the figure includes both the far upstream ($\xi\to-1$) and far downstream ($\xi\to+1$).

\section{Non-relativistic shocks in $\Dim$ dimensions}
\label{sec:non-rel}

It is interesting to generalize the classical spectrum (\ref{eq:ClassicSE3D}) of particles accelerated in a non-relativistic shock for an arbitrary dimension $\Dim\geq2$.
One way to do so is,
in analogy with \citet{Krymskii_1977}, by generalizing the steady-state Fokker--Planck equation
\citep{parker1965passage},
\begin{equation}
\label{eq:FPa}
\bm{\nab}(\np \bm{\beta}) = \frac{1}{c}\bm{\nab}\cdot \left[\mathcal{D}\bm{\nab} \np\right] + \frac{1}{\Dim}\pr_p (\np p)\bm{\nab}\cdot \bm{\beta}
\comma
\end{equation}
to $\Dim$ dimensions, where
$\mathcal{D}(\bm{x},\bm{p})$ is the spatial diffusion coefficient,
\begin{equation}
\np\equiv
p^{\Dim-1} \int f(\bm{x},\bm{p}) d\Omega
\end{equation}
is the specific (per unit momentum) number density of the accelerated particles,
and
$d\Omega=\sin^{\Dim-2}{(\theta_1)}\sin^{\Dim-3}{(\theta_2)}\ldots \sin{(\theta_{\Dim-2})}d \theta_1 d \theta_2\ldots d\theta_{\Dim-2}d\phi$ is the solid angle interval in $\Dim$ dimensions.
Here,
$\theta_i\in [0,\pi]$ are polar angles, and $\phi\in [0,2\pi)$ is an azimuthal angle.

Equation (\ref{eq:FPa}) includes, in addition to convection and diffusion terms, also a term accounting for the $p\propto \rho^{1/\Dim}$ particle momentum boost, with $\rho$ being the mass density of the plasma, due to shock compression,
\begin{align}\label{eq_adComp}
    \frac{d\np}{dt}\bigg\vert_{\text{ad}}&=\frac{d}{dt}\left[-\pr_p n(>p)\right]_{\text{ad}}=-\pr_p\left[\frac{d n(>p)}{d t}\bigg\vert_{\text{ad}}
    \right]\\
    &=-\pr_p\left[\frac{dp}{dt}\np\right]
    =\frac{1}{\Dim}\pr_p(\np p)\bm{\nabla} \cdot \bm{\beta} \nonumber\comma
\end{align}
mediated in the DSA picture by magnetic structures.
Here, $n(>p)\equiv\int_p^\infty \np \, dp$ is the number density of particles with momentum larger than $p$,
and in the last equality of Eq.~(\ref{eq_adComp}) we used the continuity equation,
\begin{equation}
    \frac{d\rho}{dt}=-c\rho \bm{\nabla}\cdot\bm{\beta} \fin
\end{equation}

Using the boundary condition of no energetic particles reaching infinitely far upstream, integration of Eq.~(\ref{eq:FPa}) yields
\begin{equation}
\np \beta=\frac{1}{c}\mathcal{D}\pr_{z} \np-\frac{\beta_u-\beta_d}{\Dim} H(z)\pr_p(\np p )|_{z=0}  \comma
\end{equation}
where $H$ is the Heaviside step function.
The solution $\np$ to this equation decays exponentially upstream and is uniform downstream, imposing the requirement
\begin{equation}
\np_d = -\frac{\myComp-1}{\Dim} \pr_p(p\np_d )|_{z=0} \fin
\end{equation}
The implied energy spectral index,
\begin{equation} \label{eq:spec_NR_2D}
\Se=1+\frac{\Dim}{\myComp-1} \comma
\end{equation}
is then a function of $\Dim$ and $\myComp$ alone.
For a strong shock in a mono-atomic ideal gas, $\myComp=1+\Dim$, and so Eq.~(\ref{eq:spec_NR_2D}) curiously yields $\Se=2$, regardless of the dimension.

It is useful, here and in anticipation of \S\ref{sec:otherMethods}, to introduce an alternative method for deriving the spectrum, by considering the fractional energy gain $g$ and return probability $P_{\text{ret}}$ of a particle undergoing a Fermi cycle \citep{Fermi49}, crossing the shock back and forth.
We thus generalize the computation of \citet{Bell_1978} to $\Dim$ dimensions,
deriving the spectral index as
\begin{equation}\label{eq:spec}
    \Se\simeq 1-\frac{\ln{(P_{\text{ret}})}}{\ln{\left\langle 1+g\right\rangle}}\fin
\end{equation}
Here, we define angular brackets
\begin{equation}
    \left\langle \ldots \right\rangle=\frac{\int \ldots\, dj}{\int dj}
\end{equation}
as averaging over the flux crossing the shock,
\begin{equation}\label{eq:fluxel}
    dj=(\mu+\beta)q(0,\mu)d\Omega\fin
\end{equation}

Consider a relativistic particle in a non-relativistic flow, crossing the shock front to the upstream region at some angle $-1\leq\mu_-\leq-\beta$, and subsequently, after
some scattering,
crossing back downstream at an angle $-\beta\leq \mu_+\leq 1$.
We choose $\mu_+$ and $\mu_-$ in the downstream frame, although this choice does not affect the energy gain in the non-relativistic shock regime.
Neglecting correlations between $\mu_-$ and $\mu_+$ (for a discussion of such correlations, see \Arad),
The flux-averaged fractional energy gain in the downstream frame is
\begin{equation}\label{eq_energyGain}
 \left\langle g \right\rangle\equiv  \left\langle \frac{E_{i+1}}{E_i}\right\rangle-1 =  \left\langle\db (\mu_+-\mu_-)\right\rangle + O(\beta^2)
\comma
\end{equation}
where $E_{i}$ is the particle energy in the $i$-th cycle and $\db\equiv(\beta_u-\beta_d)/(1-\beta_u \beta_d)\simeq(\myComp-1)\beta_d$
is the relative velocity between upstream and downstream frames.
We note that the fractional energy gain may similarly be calculated in the upstream frame, with the advantage of diminished correlations between $\mu_+$ and $\mu_-$; this is utilized in the relativistic shock analysis of \S \ref{sec:otherMethods}.

For $\Dim>1$, averaging $\mu_+$ and $\mu_-$ over the flux element of Eq.~(\ref{eq:fluxel}), and assuming an approximately isotropic PDF,
$q(0,\mu)\propto 1+O(\beta)$,
Eq.~(\ref{eq_energyGain}) yields
\begin{equation}\label{eq_3dGain}
    \left\langle g\right\rangle =  \left.\db\sqrt{\pi}\,\Gamma{\left(\frac{1+\Dim}{2}\right)} \middle/ \Gamma{\left(1+\frac{\Dim}{2}\right)}+O(\beta^2)\right.\comma
\end{equation}
where $\Gamma(x)$ is the Gamma-function.

The probability $P_\text{ret}$ of a particle crossing the shock downstream to return upstream
may be found from
the particle flux crossing the shock in each direction.
Assuming that the downstream PDF is isotropic up to second-order corrections, $q(0,\mu)\propto 1+O(\beta^2)$, we have
\begin{equation}\label{eq:pesc}
P_{\text{ret}}\equiv -\frac{j_d^-}{j_d^+}
\simeq 1- \left. 2 \beta_d \sqrt{\pi}\,  \Gamma{\left(\frac{1+\Dim}{2}\right)}\middle/\Gamma{\left(\frac{\Dim}{2}\right)} +O(\beta^2)\right.
\mbox{.}
\end{equation}
Further assuming that in the non-relativistic case $g\ll1$ and $1-P_{\text{ret}} \ll 1$, and using Eqs.~(\ref{eq_3dGain}) and (\ref{eq:pesc}), Eq.~(\ref{eq:spec}) yields
\begin{equation}
    \Se
    \simeq 1+\frac{1-P_{\text{ret}}}{\left\langle g\right\rangle}
    \simeq 1+\frac{2\beta_d \Gamma\left(1+\frac{\Dim}{2}\right)}{\beta_r \Gamma\left(\frac{\Dim}{2}\right)}=1+\frac{\Dim}{\myComp-1}\comma
\end{equation}
in agreement with Eq.~(\ref{eq:spec_NR_2D}).

It should be noted that both of these methods for deriving the spectrum implicitly assume that the scattering function that governs the particle evolution is not too anisotropic. The former method assumes that the approximation of spatial diffusion, which is not generally applicable in the vicinity of the shock, can be globally applied. The latter method assumes that the angular distribution in the downstream frame is isotropic up to corrections that are second order in $\beta$, or at least guarantee that the integral $I=\int_{-1}^1 \mu \, q(0,\mu )(1-\mu^2)^{(\Dim-3)/2}\, d\mu$ is of order $O(\beta)$ or smaller, otherwise Eq.~(\ref{eq:pesc}) has a correction term $\beta_d (1-\Dim) I$.
For a more detailed analysis of these assumptions and their implications, see \citet{KeshetEtAl19}.

\section{Relativistic shocks in $\Dim=2$ dimensions}
\label{sec:2D}

In this section we focus on shocks in $\Dim=2$ spatial dimensions.
The transport equation for small-angle scattering around an arbitrarily relativistic shock is presented in \S \ref{sec:formalism_2d}.
In \S \ref{sec:approx}, we derive a precise relationship between the spectrum and the anisotropy of shock-grazing particles, and use it to derive an analytic expression for the spectrum in the isotropic scattering limit.
We solve the problem in this limit in \S\ref{sec:eigen2}, using an expansion in analytic eigenfunctions upstream, analogous to the numeric 3D approach of \citet{Kirk_2000}. In \S\ref{sec:otherMethods}, we discuss the ultra-relativistic shock limit.

\subsection{DSA in relativistic shocks: 2D setup}
\label{sec:formalism_2d}

Assuming small-angle scattering prescribed by some velocity-angle diffusion function $\myD(z,p,\phi)$, the transport equation for
high-energy particles in two spatial dimensions is derived in Appendix \S\ref{sec:DiffusionEq}.
For a steady state developing in the shock frame, one finds
\begin{equation} \label{eq_transport1}
c \gamma (\mu+\beta) \pr_{z} f(z,p,\phi) = \pr_\phi \left(\myD \pr_\phi f \right) \fin
\end{equation}

Boundary conditions include continuity across the shock front,
$f_u(z=0,p_u,\phi_u)=f_d(z=0,p_d,\phi_d)$, and no flux reaching far
upstream, $f_u(z\to-\infty)=0$. Here, upstream and downstream quantities
are related by a Lorentz boost of velocity $\beta_r$,
$p_d=\gamma_r p_u(1+\beta_r \mu_u)$ and $\mu_d=(\mu_u+\beta_r)/(1+\beta_r \mu_u)$.

Imposing the implied power-law spectrum,
$f=\myQ(z,\phi)p^{-\SP}$, and assuming that $\myD$ is separable in the form $D_2(z,p)D(\phi)$, equation (\ref{eq_transport1})
 becomes
\begin{equation} \label{eq_transport2}
(\mu+\beta) \pr_\tau \myQ(\tau,\phi) = \pr_{\phi}\left[D(\phi)  \partial_{\phi}\myQ\right]
\comma
\end{equation}
where we defined $\tau\equiv (c\gamma)^{-1}\int_0^{z} D_2(\check{z},p)d\check{z}$ as the shock-frame optical depth.
In the case of isotropic diffusion, $D=\const$, Eq.~(\ref{eq_transport2}) can be solved by separation of variables
\citep[\eg][]{Kirk_2000}.
The angular functions in 2D are then given by the periodic Mathieu functions, also known as elliptic cosine functions, utilized in \S\ref{sec:eigen2}.

\subsection{Analytic spectrum--anisotropy connection
}
\label{sec:approx}

Following {\KW} and using similar notations, we exploit the stationary nature of the PDF at shock-grazing angles, where $\mu+\beta=0$.
We expand the PDF and diffusion function in each frame about the grazing angle,
\begin{equation}
 \myQ(\tau,\phi)=a_0(\tau)+a_1(\tau)(\mu+\beta)+a_2(\tau)(\mu+\beta)^2+\ldots
\end{equation}
and
\begin{equation}
 D(\phi)=d_0+d_1(\mu+\beta)+d_2(\mu+\beta)^2+ \ldots
\end{equation}
The transport equation (\ref{eq_transport2}) then implies the precise relation
\begin{align}\label{eq:diffusion_coeffs}
    \frac{2 a_2}{a_1}+\frac{d_1}{d_0}+\beta \gamma^2=0 \fin
\end{align}
Notice that here we assumed that the PDF is an analytic function near the grazing angle.

Next, we use continuity across the shock to relate the upstream and downstream expansions of \myQ. To first order, this yields
\begin{equation}\label{eq:s_vs_a}
    r_u+\SP\beta_u=r_d+\SP\beta_d \comma
\end{equation}
where
\begin{equation}
r\equiv \gamma^{-2}\frac{a_1(\tau=0)}{a_0(\tau=0)}
\end{equation}
is a measure of the PDF anisotropy along the shock.
Using Eq.~(\ref{eq:diffusion_coeffs}), the second-order expansion of continuity across the shock then provides a precise relation between the particle spectrum and the anisotropy along the shock front,
\begin{equation}\label{eq:s_vs_r_2D_Assaf}
   \SP(\SP+1)\left[ \beta^2 \right]+(2\SP+1) \left[ r\beta \right] = \left[r d \right] \comma
\end{equation}
where we defined square brackets $\left[ \ldots \right] \equiv (\ldots)_u-(\ldots)_d$ as an operator taking the difference across the shock.
Here, $d\equiv\gamma ^{-2}d_1/d_0$ is a measure of the diffusion function anisotropy near the grazing angle.
For isotropic diffusion, $d=0$, and the RHS of the equation vanishes.

Combining Eqs.~(\ref{eq:s_vs_a}) and (\ref{eq:s_vs_r_2D_Assaf}), we may now derive $\SP$ as a function of the anisotropy parameter $r$ in any frame. For example, in terms of $r_u$,
\begin{equation}\label{eq:s_vs_qu}
    \SP=\frac{r_u+\frac{\beta_u+d_d}{2} \pm \sqrt{\left(r_u+\frac{\beta_u+d_d}{2}\right)^2+r_u[d-\beta]}}{-(\beta_u-\beta_d)}\fin
\end{equation}
For isotropic diffusion, $d=0$, and Eq.~(\ref{eq:s_vs_qu}) simplifies to
\begin{equation}\label{eq:s_iso_ru}
   \SpIso=-\frac{r_u+\frac{\beta_u}{2}\pm\sqrt{r_u^2+\beta _d r_u+\frac{\beta _u^2}{4}}}{\beta_u-\beta_d}\fin
\end{equation}
Analogous expressions for $\SP$ as a function of $r_d$ are obtained by interchanging subscripts $u\leftrightarrow d$ and reversing the sign of the $[\ldots]$ operator in Eqs.~(\ref{eq:s_vs_qu}) and (\ref{eq:s_iso_ru}).

One can also express the spectrum as a function of the shock-frame grazing anisotropy, which we define as $r_s\equiv a_1^{(s)}/a_0^{(s)}$.
Expanding $\myQ$ around the shock-frame grazing angle, $\mu_s=0$, and using continuity across the shock to relate $r_s$ and, say, $r_u$, one finds
\begin{equation}
    r_s=r_u+\SP\beta_u \fin
\end{equation}
Plugging this result into Eqs.~(\ref{eq:s_vs_a}) and (\ref{eq:s_vs_r_2D_Assaf}) yields
\begin{equation}\label{eq:s_vs_qs}
    \SP=
     \frac{ r_s+\frac{[\beta d]}{2[\beta]}
    \pm \sqrt{\left( r_s +\frac{[\beta d]}{2[\beta]} \right)^2+ r_s b\left(1- \frac{[d]}{[\beta]}\right) } }{b}\comma
\end{equation}
where $b\equiv\beta_u+\beta_d$.
For isotropic diffusion, Eq.~(\ref{eq:s_vs_qs}) simplifies to
\begin{equation}
     \SpIso=\frac{r_s}{\beta_u+\beta_d}\left(1 \pm \sqrt{1+\frac{\beta_u+\beta_d}{r_s}}\right)\fin
\end{equation}

Equations~(\ref{eq:s_vs_qu}) and (\ref{eq:s_vs_qs}) provide a powerful, precise connection
between the spectrum and the anisotropy of shock-grazing particles.
These results do not rely on the test-particle approximation, and are valid for any small-angle scattering described by $D$.
The 3D analogue of this spectrum--anisotropy connection was derived in \KW.

The spectrum--anisotropy relation may be used to estimate the spectrum, if one can constrain the grazing anisotropy parameter $r$.
One useful constraint arises from the limit of infinite compressibility, $\Ad\to 1$, where
$\beta_d=0$.
Here, the escape probability vanishes, so Eq.~(\ref{eq:spec}) implies a spectral index $\Se\to 1$ ({\KW}).
Another constraint is obtained in the non-relativistic shock limit, studied in \S\ref{sec:non-rel}.
In 2D, Eq.~(\ref{eq:spec_NR_2D}) yields $\Se=(\myComp+1)/(\myComp-1)$ in this limit, so in a strong non-relativistic shock in 2D one may infer, for example, that $r_d=[2\beta_u(2\beta_u-\beta_d-d_u)]/(5\beta_u-\beta_d+d_d-d_u)$.

Next, we derive an expression for the spectrum in the special case of isotropic diffusion, following the method of {\KW}.
We focus on the downstream frame, where, unlike in the upstream, the anisotropy does not become very strong even in the ultra-relativistic shock limit.
Combining Eqs.~(\ref{eq:spec_NR_2D}), (\ref{eq:s_vs_a}) and (\ref{eq:s_vs_r_2D_Assaf}), we rewrite Eqs.~(\ref{eq:s_vs_a}-\ref{eq:s_vs_r_2D_Assaf}) as
\begin{equation} \label{eq:s_vs_r_2D_iso}
r_d = \frac{s_p^2(\beta_u-\beta_d)-\SP\beta_d}{2\SP+1} \fin
\end{equation}
In the non-relativistic shock limit, we may now quantify the downstream anisotropy as
\begin{equation}\label{eq:a1toa0}
\frac{a_1^{(d)}}{a_0^{(d)}} = \frac{2\beta_u(2\beta_u-\beta_d)}{5\beta_u-\beta_d} \fin
\end{equation}

Equation (\ref{eq:a1toa0}) holds not only in the non-relativistic shock limit, but also for any $\beta_u$ in the infinite compressibility limit where $\beta_d=0$.
Following \KW, we extrapolate the result for arbitrary $\{\beta_u,\beta_d\}$; the reasoning for this extrapolation is further discussed in \S\ref{sec:Summary}.
Plugging Eq.~(\ref{eq:a1toa0}) into the downstream version of Eq.~(\ref{eq:s_iso_ru}) finally yields
\begin{align} \label{eq:s_iso_2D}
\SpIso  & \to  \frac{r_d+\frac{\beta_d}{2}+\sqrt{r_d^{2}+r_d
\beta_u+\frac{\beta_d^2}{4}}}{\beta_u-\beta_d} \\
& = \frac{\frac{1}{2}+\frac{2\myComp}{\gamma_d^2} \frac{2\myComp-1}{5\myComp-1}+\sqrt{\frac{1}{4} + \frac{2\myComp^2}{\gamma_d^2} \frac{2\myComp-1}{5\myComp-1} + \frac{4\myComp^2}{\gamma_d^4} \left(\frac{2\myComp-1}{5\myComp-1}\right)^2 }}{\myComp-1} \fin
\nonumber
\end{align}

\begin{figure}[htb]
\centerline{\epsfxsize=9cm \epsfbox{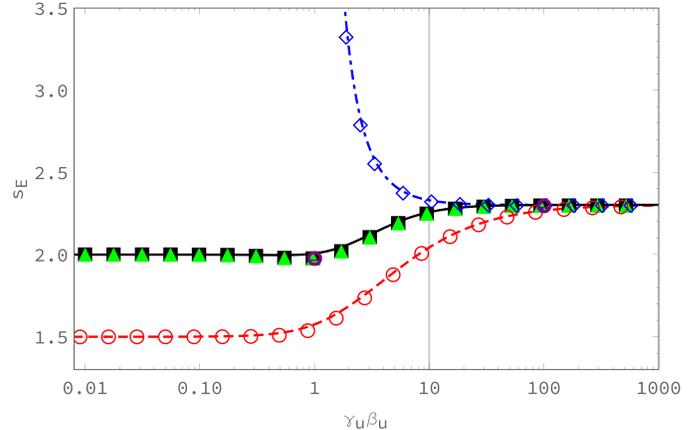}}
\caption{
The energy spectral index $\Se$ for isotropic diffusion as a function of the shock covariant velocity $\gamma_u\beta_u$,
computed with the upstream eigenfunction expansion
(symbols, with $M=6$ eigenfunctions) and with the analytic approximation (\ref{eq:s_iso_2D}) (curves),
shown for three different EOS: the JS EOS (black squares and a solid curve), an adiabatic index $3/2$ (red circles and dashed curve) and for a relativistic gas where $\beta_u\beta_d=1/2$ (blue diamonds and dot-dashed curve).
For $\gamma_u\beta_u\geq10$ (to the right of the vertical dotted-line), the eigenfunctions and eigenvalues are calculated using the ultra-relativistic approximation (\ref{eq_ceApprox}).
Convergence is demonstrated using $M=25$ moments (purple circles with error bars; see \S\ref{sec:err}).
The first upstream eigenfunction alone provides an excellent approximation, as shown using Eq.~(\ref{eq_zero-overlap}) for the J\"{u}ttner--Synge case (green triangles, using the exact eigenfunctions).
}
\label{fig:p_index_2D}
\end{figure}

In resemblance of the 3D case, we find that also in 2D, the result (\ref{eq:s_iso_2D}) of the above extrapolation is consistent with the spectrum computed in other, numerical or semi-analytical methods, for an arbitrarily relativistic shock and any EOS.
Figure \ref{fig:p_index_2D} shows that our analytic estimate is in excellent agreement with the upstream eigenfunction expansion presented in \S \ref{sec:eigen2}, in both non-relativistic and ultra-relativistic limits.
In particular, in the ultra-relativistic shock limit, $\beta_u\to1$ and $\beta_d\to1/2$, Eq.~(\ref{eq:s_iso_2D})
implies that
\begin{equation} \label{eq:sE_Urel_2D}
\SpIso(\gamma_u\to\infty) \to \frac{3+\sqrt{13}}{2} \simeq 3.303 \comma
\end{equation}
consistent within $0.2\%$ with the eigenfunction method.
Interestingly, in the trans-relativistic regime, $\gamma \beta \approx 1$, there is a slight, $\sim 1\%$ deviation between the two methods.

\subsection{Upstream eigenfunction expansion}
\label{sec:eigen2}

Next, we focus on the case of isotropic diffusion, $D=\const.$, where the problem becomes largely analytically tractable.
Here, the transport equation (\ref{eq_transport2}) can be directly solved by expanding the PDF in upstream eigenfunctions, in a method parallel to that applied by \citet{Kirk_2000} for the three-dimensional case. An advantage of working in two-dimensions is that the eigenfunctions reduce to the well-known elliptic cosine functions, as we show below.

Separating the PDF variables, let
\begin{equation}
\myQ(\tau,\phi)\equiv T(\tau)\Phi(\phi)\fin
\end{equation}
Plugging this into Eq.~(\ref{eq_transport2}), we obtain two separate equations, connected by an eigenvalue $\myq$ which we define such that
\begin{equation} \label{eq_TauDiff}
T'(\tau)=\frac{\myq}{2} T(\tau)
\end{equation}
and
\begin{equation} \label{eq_PhiDiff}
\Phi''(\phi) = \frac{\myq}{2} (\beta +\mu)\Phi(\phi) \fin
\end{equation}
The spatial equation (\ref{eq_TauDiff}) indicates an exponential spatial dependence,
\begin{equation}
T(\tau) \propto  e^{\frac{\myq}{2} \tau} \comma
\end{equation}
where the boundary conditions dictate that $\Lambda>0$ upstream and $\Lambda\leq0$ downstream.

The solution to the angular equation (\ref{eq_PhiDiff}) under our assumption of axisymmetry, $q(\tau,-\phi)=q(\tau,\phi)$,
is given by
\begin{equation}
\label{PhiEvenSolution}
\Phi(\phi) \propto  \text{C}\left(-2\beta\myq,\myq,\frac{\phi}{2}\right)\comma
\end{equation}
where
$\mbox{C}(a,\Lambda,x)$ are the Mathieu cosine functions \citep[see, \eg][]{mclachlan1951theory},
defined as the solutions of the Mathieu equation
\begin{equation}
   \frac{d^2 \text{C}}{dx^2}+\left[a-2\Lambda \cos{(2x)}\right]\text{C}=0
\end{equation}
that are even in $x$, namely $\text{C}(a,\Lambda,x)=\text{C}(a,\Lambda,-x)$.

Next, we impose $2\pi$-periodicity in $\phi$, which corresponds to $1\pi$-periodicity in $x$.
In general, for a given $\myq$, $\text{C}(a,\myq,x)$
becomes periodic in $x$ only for a discrete, infinite set of so-called characteristic values $a=a_r(\myq)$, which are
the roots of a continued fraction equation \citep[][]{ince1927mathieu}, which we write as
\begin{equation}
    \frac{1}{(4-a)\Lambda^{-2}+4a^{-1}}= \myK_{k=1}^\infty\frac{-\myq^2}{4k^2-a} \fin
\end{equation}
Mathieu functions with $1\pi$-periodicity correspond to $a_r(\myq)$ characteristic values with an even index, $r=2j$, where $j=\{0,1,2\ldots\}\geq 0$. Here, $r$ is an index, not to be confused with the anisotropy parameter (which we defined in \S\ref{sec:approx} and do not use in the current section).

In Eq.~(\ref{PhiEvenSolution}), $a=-2\beta\Lambda=a_r(\myq)$, which we can now solve for $\Lambda$.
For each $j$ we find two such solutions, denoted $\myq_{\pm j}(\beta)$, namely
\begin{equation}\label{eq_eigenvalue}
    -2\beta \myq_{\pm j} = a_{2j}(\myq_{\pm j}) \fin
\end{equation}
For $j>0$, the solutions satisfy $\myq_j>0$ and $\myq_{-j}<0$.
For $j=0$, there is one positive solution, denoted $\myq_{0_+}$, and one trivial solution, denoted $\myq_{0_-}=0$.
This behavior is guaranteed by the Sturm-Liouville theory, noting that the even-parity function $\text{C}(a_{2j},\myq,x)$ have $2j$ zeros in the interval $0<x<\pi$ \citep[][section 2.152]{mclachlan1951theory}.

The Mathieu functions $\text{C}(a_r,\myq,x)$ that have even parity are known as the elliptic cosine functions \citep[sometimes referred to as cosine elliptic functions; see \eg][\S 2.13]{mclachlan1951theory}, and are denoted $\text{ce}_{r}(x,\myq)$.
Our $1\pi$-periodic functions can therefore be written as
\begin{equation}
\Phi_{\pm j}(\phi) = \text{ce}_{2j}\left(\frac{\phi}{2},\myq_{\pm j}\right) \fin
\end{equation}
The eigenfunctions obey the orthogonality relation
\begin{equation}\label{eq_MathieuOrth}
\int\limits_{0}^{\pi}(\mu+\beta)\Phi_i(\phi)\Phi_j(\phi)d\phi=w_i(\beta)\delta_{ij}\comma
\end{equation}
where $w_i(\beta)$ are constant weights.
This result can be verified by
comparing Eq.~(\ref{eq_MathieuOrth}) after applying Eq.~(\ref{eq_PhiDiff}) to either $\Phi_i$ or $\Phi_j$, and integrating by parts.
Limiting expressions for all eigenvalues and eigenfunctions in the $\gamma\gg1$ limit are discussed in \S \ref{sec:otherMethods}.
In the limit $\Lambda\to 0$, $\Phi_{\pm j}$ reduces to $\cos(2j\phi)$.

The upstream PDF can now be written in the form
\begin{equation} \label{eq_DistributionUp}
f_u=p_u^{-\SP}\sum_{i=0_+}^{\infty} \kappa_i \, \text{ce}_{2i}\left(\frac{\phi_u}{2},\myq_i\right) e^{\myq_i\tau/2} \comma
\end{equation}
where $\kappa_i$ are constants.
An analogous expansion can be carried out downstream, if necessary.
As $f_d$ is not composed of any $j\geq0_+$ eigenfunctions, the orthogonality relation (\ref{eq_MathieuOrth}) implies that
\begin{equation} \label{eq:ShockOrtho}
\int\limits_{0}^{\pi}(\mu_d+\beta_d)f_d(\tau=0)\Phi_{d,j}(\phi_d)d\phi_d = 0 \quad \forall j\geq 0_+ \fin
\end{equation}

To derive the spectrum, we approximate $f_u$ by truncating the sum in Eq.~(\ref{eq_DistributionUp}) as $0_+\leq i\leq\Nfs-1$, thus using only $\Nfs$ upstream eigenfunctions, with $\Nfs$ unknown coefficients $\kappa_i$.
The spectrum and $\kappa_i$ can be determined by examining $\Nfs$ out of the equations (\ref{eq:ShockOrtho}), conveniently chosen with $0_+\leq j\leq \Nfs-1$.
These $\Nfs$ equations can be compactly written as
\begin{equation}\label{eq_sMatrix}
\sum_{j=0_+}^{\Nfs-1}S_{ij}\kappa_{j}=0
\comma
\end{equation}
where the $S_{ij}$ matrix elements are proportional to the downstream-frame overlap integral of the upstream $i$ and downstream $j$ eigenfunctions,
\begin{equation}
\label{eq_sMatrixValues}
\hspace{-0.2cm} S_{ij} \equiv \int\limits_{0}^{\pi}(\beta_d+\mu_d)(1+\beta_r \mu_u)^\SP \Phi_{u,i}(\phi_u)\Phi_{d,j}(\phi_d)d\phi_d \comma
\end{equation}
which we compute numerically for each element.
Equation (\ref{eq_sMatrix}) has a non-trivial solution only when $|S_{ij}|=0$, setting the condition for finding the spectral index. The approximate PDF is then found using the null space of $S_{ij}$ as the upstream coefficients $\kappa_i$.

We use this method to calculate the spectral index for three different EOS: the JS EOS, which describes an arbitrarily relativistic ideal gas; a fixed adiabatic index $\Ad=3/2$; and a relativistic gas where $\beta_u\beta_d=1/2$.
The energy spectral index $\Se$ resulting from this upstream eigenfunction expansion is shown (symbols) in Figure \ref{fig:p_index_2D}, and found to be in very good agreement with the analytic approximation of \S\ref{sec:approx} (curves).
The figure uses $M=6$ upstream eigenfunctions, but we demonstrate the convergence (error bars) by varying $M$ up to 25, as well as the details of the numerical integration of Eq.~(\ref{eq_sMatrixValues}); for details, see  Appendix \S \ref{sec:err}.

\begin{figure*}
\centerline{\epsfxsize=18.5cm \epsfbox{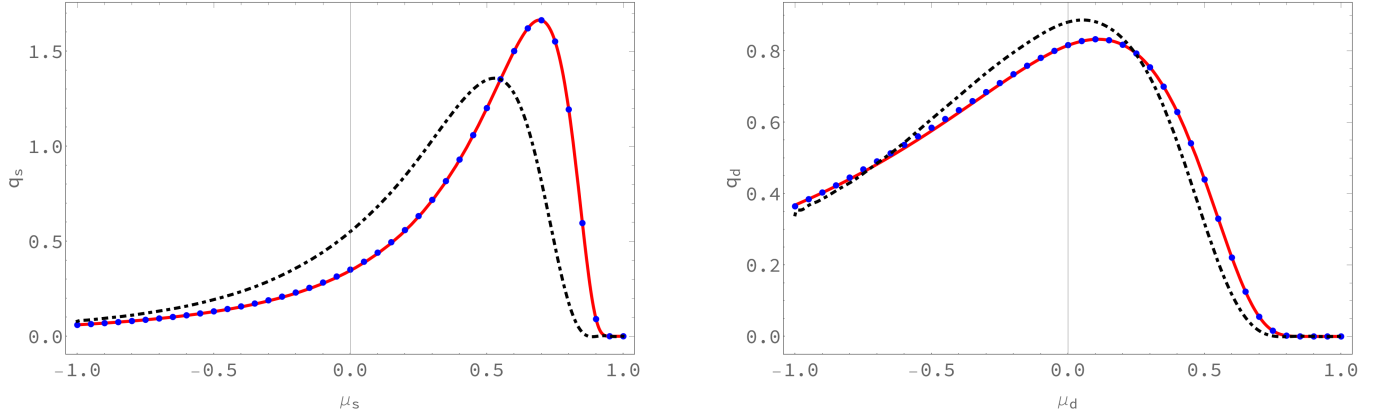}}
\caption{
Shock-front PDF in both shock (left panel) and downstream (right) frames, for a strong shock with $\gamma_u \beta_u=10$, the JS EOS, and isotropic scattering.
The angular PDF is shown based on an $M=10$ upstream eigenfunctions expansion (blue circles), and on the first eigenfunction (solid red line), and is compared to the 3D PDF computed with the moment expansion method \citep[][and A20, dot-dashed black line]{Keshet06}.
The PDF are normalized such that the integral over the respective $\mu$ is unity. }
\label{fig:distribution_2D}
\end{figure*}

Figure \ref{fig:distribution_2D} presents the angular PDF computed for $\gamma_u\beta_u=10$ with the JS EOS,
in both shock and downstream frames.
The PDF in 2D behaves qualitatively similarly to the 3D PDF \citep{Kirk_2000}, showing almost no particles crossing the shock from upstream to downstream along the shock normal, and an attenuated flux of particles crossing in the opposite direction.
As the figure illustrates, the first upstream eigenfunction provides a remarkably good approximation to the full PDF,
\begin{equation}\label{eq_FirstEig}
\myQ_u(0,\phi_u) \simeq \Phi_{u,0_+}(\phi_u) \propto  \text{ce}_{0_+}\left(\frac{\phi_u}{2},\myq_{0_+}^{(u)}\right)\comma
\end{equation}
again in resemblance of the 3D case \citep[][]{Kirk_2000}.
The first eigenfunction remains an excellent approximation for an arbitrarily relativistic shock and any EOS.

\begin{figure}[h!]
\centerline{\epsfxsize=8.5cm \epsfbox{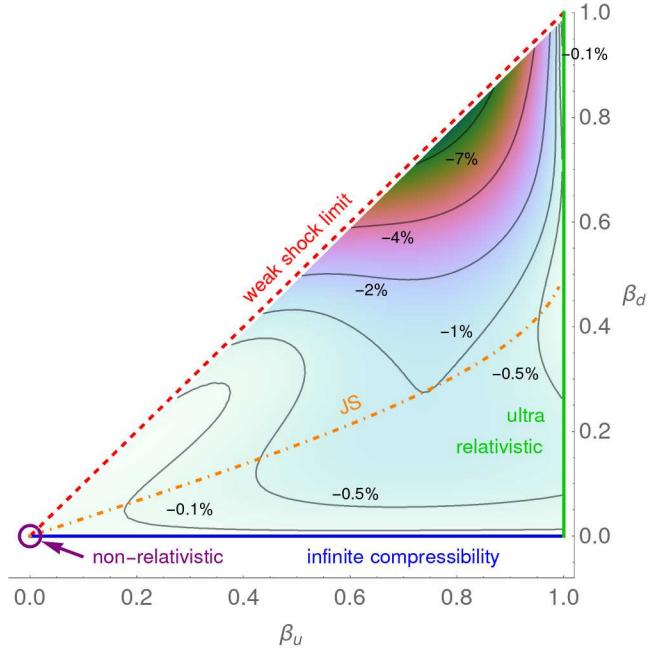}}
\caption{
Normalized difference $s_p^{(M=1)}/s_p-1$ (contours) between the spectrum inferred from the first-eigenfunction overlap (\ref{eq_zero-overlap}) and from the analytic expression (\ref{eq:s_iso_2D}), for isotropic scattering in 2D, shown in the full $\{\beta_u,\beta_d\}$ phase space (including non-physically large $\beta_d$).
Also shown are the shock jump conditions for the JS EOS (dot-dashed green curve), and limiting cases (labeled lines and circle) where Eq.~(\ref{eq:a1toa0}) can be accurately or at least reasonably applied (see \S\ref{sec:Summary}).
}
\label{fig:PhaseSpace}
\end{figure}

Moreover, using only the first eigenfunction upstream and the first eigenfunction downstream, namely choosing $M=1$, is sufficient to obtain an accurate approximation for the spectrum, as shown in Figures \ref{fig:p_index_2D} and \ref{fig:PhaseSpace}.
Here, $\SP$ is determined by requiring zero overlap between the first eigenfunctions in each frame, namely by solving the integral equation
\begin{align} \label{eq_zero-overlap}
   0= \int\limits_{0}^{\pi} & (\beta_d+\mu_d)(1+\beta_r \mu_u)^\SP \\
  & \times \text{ce}_{0_+}\left(\frac{\phi_u}{2},\myq_{0_+}^{(u)}\right) \text{ce}_{0_+}\left(\frac{\phi_d}{2},\myq_{0_+}^{(d)}\right)d\phi_d \fin \nonumber
\end{align}
In the ultra-relativistic limit, the first eigenfunction in Eq.~(\ref{eq_zero-overlap}) alone yields
\begin{equation}
\Se=2.2988 \pm 0.0001 \comma
\end{equation}
obtained by extrapolation to $\gamma_u\beta_u\to\infty$.

\subsection{Ultra-relativistic limit}\label{sec:otherMethods}

In the ultra-relativistic limit, the upstream frame eigenvalues become large, $\myq\gg 1$, and limiting expressions of all eigenvalues and eigenfunctions may be found in \citet[][]{ogilvie2015rigorous}.
Simplifying their expressions, in this limit the upstream eigenvalues
are given by
\begin{equation}\label{eq_lambdaApprox}
  \myq_j=4\gamma^4(1+4j)^2+O(\gamma^2)\comma
\end{equation}
and the upstream eigenfunctions become, in terms of $y\equiv (1+\mu)\gamma^2$,
\begin{align}\label{eq_ceApprox}
    \text{ce}_{2j}(y)&\simeq e^{-\frac{\sqrt{\Lambda_j}}{2\gamma^2}y} \, _1F_1\left(-j;\frac{1}{2};\frac{\sqrt{\Lambda_j}}{\gamma^2} y \right)=e^{-\frac{\sqrt{\Lambda_j}}{2\gamma^2}y} \\
    &\times\left[1-j+\sum _{k=0}^j j \left(-\frac{2\sqrt{\Lambda_j}}{\gamma^2}y \right)^k \prod _{i=2}^k \frac{j-i+1}{i (2 i-1)}\right]\nonumber\comma
\end{align}
up to an inconsequential normalization.
In particular, using the first eigenfunction only, we may approximate
\begin{equation}\label{eq_eigApp1}
\myQ_u(0,\phi_u)\propto  \text{ce}_{0_+}\left(\frac{\phi_u}{2},\myq_{0_+}^{(u)}\simeq 4\gamma_u^4\right) \appropto e^{-\left(1+\mu_u \right)\gamma^2}\fin
\end{equation}

The approximation (\ref{eq_ceApprox}) may also be derived by an asymptotic analysis of the angular transport equation (\ref{eq_PhiDiff}) in the upstream frame \citep[in resemblance of][but note the factor $2$ difference between our definitions of $\{\Lambda,y\}$ and theirs]{kirk1989particle}. After a change of variables, Eq.~(\ref{eq_PhiDiff}) becomes
\begin{equation}
    2y\left(1-\frac{y}{2 \gamma^2}\right)\Phi''(y)+\left(1-\frac{y}{\gamma^2}\right)\Phi'(y)=\frac{\myq}{\gamma^4}(2y-1)\Phi(y) \fin
\end{equation}
Taking into account that in the region $\mu>-\beta$ the eigenfunctions are exponentially small, we focus on the regime $-1\leq\mu\leq-\beta$.
Noting that $y/\gamma^2=1+\mu\ll 1$, we neglect such terms. Additionally, we  plug in Eq.~(\ref{eq_lambdaApprox}), to yield
\begin{equation}\label{eq_eigenApprox}
      2y\Phi''(y)+\Phi'(y)=(1+4 j)^2 (2y-1)\Phi(y)\fin
\end{equation}
Bounded solutions to Eq.~(\ref{eq_eigenApprox}) are given by
\begin{equation}
   \Phi_j(y)=  e^{-y(1+4j)}\sqrt{y}\  U\left[\frac{1}{2} (1-2 j),\frac{3}{2},2 (4 j+1) y\right]\comma
\end{equation}
where $U(a,b,z)$ is the confluent hypergeometric function of the second kind. For integer $j\geq 0$, this is equivalent to Eq.~(\ref{eq_ceApprox}) up to a normalization.

The above approximations are useful because it is exceedingly difficult to compute the exact eigenfunctions as one approaches $\beta\to 1$.
Indeed, these approximations are used for the highly relativistic shocks shown in Figure \ref{fig:p_index_2D} (right of the vertical line).
For $\gamma_u\beta_u \rightarrow \infty$, we obtain the asymptotic spectrum
\begin{equation}
\label{eq.Asy_p}
\Se=2.2985\pm 0.0001\comma
\end{equation}
estimated by extrapolating the data in the figure to $\gamma_u\beta_u\to\infty$, with weights given by the convergence tests.
If we use only the first eigenfunction, this approximation (\ref{eq_eigApp1}) with the single-eigenfunction overlap (\ref{eq_zero-overlap}) yields
\begin{equation}\label{eq.Asy_pApprox}
\Se=2.2988\pm0.0001 \fin
\end{equation}

One can use the approximate first eigenfunction upstream to infer the spectrum in the ultra-relativistic shock limit, even without computing the precise elliptic cosine functions.
This can be carried out using the single-eigenfunction overlap (\ref{eq_zero-overlap}), if the first eigenfunction downstream $\Phi_{0+}^{(d)}$ can also be approximated.
As this function is positive definite, and the overlap region is near $\mu=-1$, we may approximate
\begin{equation} \label{eq:Phi0dApprox}
\Phi_{0+}^{(d)}\propto \exp[-c_1(1+\mu)^1-c_2(1+\mu)^2-\ldots-c_n(1+\mu)^n]
\end{equation}
with finite $n$ terms. Plugging this into the transport equation (\ref{eq_PhiDiff}) fixes the coefficients $c_1=(1-\beta_d)\Lambda/2$, $c_2=[(1-\beta_d)^2\Lambda-\beta_d]\Lambda/12$, etc.; the value of $\Lambda$ is then fixed by orthogonality with $\Phi_{0-}^{(d)}$, \ie
\begin{equation}
\int_0^\pi (\mu_d+\beta_d)\Phi_{0+}^{(d)}d\phi_d = 0 \fin
\end{equation}
This procedure converges rapidly; taking $n=2$ already gives $\Se\simeq 2.38$.

A simpler but less accurate method is to compute the return probability and the mean energy gain directly from the approximate first eigenfunction upstream, and then use Eq.~(\ref{eq:spec}) to determine the spectrum.
With the approximation (\ref{eq_eigApp1}), the energy gain, best computed in the upstream frame (A20), is
\begin{equation}\label{eq:energain}
    \left\langle 1+g\right\rangle = \frac{\int\frac{1+\beta_r \mu_+}{1+\beta_r\mu_-}dj_u^- dj_u^+}{\int dj_u^- dj_u^+} \simeq 2.31 \comma
\end{equation}
where index '$+$' indicates forward (\ie towards downstream) directions, $-\beta\leq\mu_+\leq1$, and index '$-$' indicates backward (toward upstream) directions, $-1\leq\mu_-\leq -\beta$.
The return probability is given by
\begin{equation}\label{eq:Pret}
P_{\text{ret}}= -\frac{\int dj_d^-}{\int dj_d^+}
\simeq 0.33 \comma
\end{equation}
where we used  $\Se= 2.30$ to obtain a numerical estimate.
If, instead, we leave $\Se$ undetermined, Eq.~(\ref{eq:spec}), yields a rather crude approximation, $\Se\simeq 2.07$.

\section{Relativistic shocks in $\Dim\geq3$ dimensions}
\label{sec:RelND}

The preceding analysis can be generalized for arbitrary $\Dim\geq 3$ dimensions.
The same assumptions leading to Eq.~(\ref{eq_transport1}) in 2D, yield the $\Dim\geq 3$ transport equation
\begin{equation} \label{eq:transportHighD}
    (\mu+\beta)\partial_\tau q(\tau,\mu)=\frac{\partial_\mu \left[(1-\mu^2)^{\frac{\Dim-1}{2}}D(\mu)\partial_\mu q\right]}{(1-\mu^2)^{\frac{\Dim-3}{2}}} \fin
\end{equation}
The spectrum--anisotropy connection (\ref{eq:s_vs_qu}) generalizes to
\begin{equation} \label{eq:GrazingND}
   \hspace{-0.3cm} \SP=
   \frac{v + \sqrt{v^2 - r_d\left[d_d-d_u-(\Dim-3)(\beta_u-\beta_d)\right]}}{\beta_u-\beta_d}
   \comma
\end{equation}
where we defined $v\equiv [(\Dim-2)\beta_u+\beta_d+d_u+2r_d]/2$, and used the downstream anisotropy parameter $r_d$ as defined in \S\ref{sec:approx}.

For isotropic diffusion, the downstream anisotropy of a non-relativistic shock is therefore given, for any $\Dim\geq2$, by
\begin{equation}\label{eq:a1toa0N}
\frac{a_1^{(d)}}{a_0^{(d)}} = \frac{\Dim \beta_u(2\beta_u-\beta_d)}{(\Dim+3)\beta_u+(\Dim-3)\beta_d} \fin
\end{equation}
Invoking the same assumptions leading to the spectrum (\ref{eq:s_iso_2D}) now gives
\begin{equation} \label{eq:GrazingNDiso}
    \SeIso = \frac{\varv+\sqrt{\varv^2+(\myComp-1)[w(\Dim+1)+(\Dim-\myComp)(\Dim-1)]}}{\myComp-1}
\end{equation}
for any $\Dim\geq2$.
Here, $\varv\equiv \Dim+w-(\Dim \myComp+1)/2$ and $w\equiv \gamma_d^{-2}\Dim\myComp(2\myComp-1)/[(\Dim+3)\myComp+\Dim-3]$.
Interestingly, as $\Dim$ becomes large, the spectrum approaches $\Se\to 1+\Dim/\myComp+O(\Dim^{-1})$, asymptotically giving the familiar $\Se\to 2$ for an arbitrarily relativistic, strong shock.

The angular component of the transport equation becomes, for $\Dim\geq 3$,
\begin{equation}\label{eq:transport3d}
(1-\mu^2)\Phi''(\mu)-(\Dim-1)\mu\Phi' = (\mu+\beta)\frac{\Lambda}{2} \Phi \fin
\end{equation}
In the upstream of an ultra-relativistic shock, we may approximate Eq.~(\ref{eq:transport3d}) around $\mu=-1$ (as in
\S\ref{sec:otherMethods}) to find
\begin{equation}\label{eq:eigenNd}
    \Phi_j(y)\simeq e^{-(\Dim-1+4 j)y}{}_1 F_1\left[-j,\frac{\Dim-1}{2},2(\Dim-1+4 j) y\right]\comma
\end{equation}
with the eigenvalues
\begin{equation}
    \Lambda_j=4\gamma^4 (4j+\Dim-1)^2+O(\gamma^2)\comma
\end{equation}
derived by assuming that
\begin{equation}
   \Phi_j=e^{-\frac{y \sqrt{\Lambda_j}}{2\gamma^2}}\sum_{n=0}^j a_n y^n
\end{equation}
\citep[\emph{c.f.}][equation A5; notice the factor two difference in the definition of $\Lambda$.]{kirk1989particle}.
The first upstream eigenfunction is therefore
\begin{equation} \label{eq:Phi0ApproxND}
\Phi_0\simeq e^{-(\Dim-1)(1+\mu)\gamma^2}
\end{equation}
for an ultra-relativistic shock in any $\Dim\geq 2$.

Given the first upstream eigenfunction, one can directly estimate the spectrum in the methods outlined in \S\ref{sec:otherMethods}, as we demonstrate for $\Dim=3$.
Approximating the first downstream eigenfunction as in Eq.~(\ref{eq:Phi0dApprox}), we rapidly converge on $\Se\simeq 2.23$; taking a single term ($n=1$) in this equation already yields $\Se\simeq2.29$.
One could derive the spectrum from the first eigenfunction using Eq.~(\ref{eq:spec}), instead.
For $\Dim=3$, the energy gain in Eq.~(\ref{eq:energain}) can be computed analytically, $\langle 1+g \rangle = 5\{1-e+3e^3[E_i(-3)-E_i(-2)]\}\simeq 2.21$, where $E_i$ is the exponential integral.
The return probability can be computed as in Eq.~(\ref{eq:Pret}), giving
$P_{\text{ret}}=1+[\Gamma(\Se+1,2)-3\Gamma(\Se,2)]/[3\Gamma(\Se,3)-\Gamma(\Se+1,3)]\simeq 0.38$, the latter estimate obtained by assuming $\Se=2.23$. Alternatively, solving Eq.~(\ref{eq:Pret}) for $\Se$ gives the approximate $\Se\simeq 2.26$.

\section{Summary and Discussion}
\label{sec:Summary}

We generalize the analysis of DSA in collisionless shocks to an arbitrary number $\Dim$ of spatial dimensions, in order to facilitate the understanding of shock studies in 2D, and in search for insights into the 3D case.
The problem, illustrated in Figure \ref{fig:distribution_map}, is solved in the test-particle, small-angle scattering approximation, first for non-relativistic shocks (\S\ref{sec:non-rel}), and with additional assumptions, also for relativistic shocks (in \S\ref{sec:2D} and \S\ref{sec:RelND}).

Curiously, for any $\Dim$, we recover the familiar, flat spectral index $\Se=2$ (see Eq.~\ref{eq:spec_NR_2D}), in which energy diverges only logarithmically, for a non-relativistic shock in a mono-atomic gas.
The same result is obtained in the $\Dim\to\infty$ limit also for a relativistic strong shock, at least when scattering is isotropic.
These results highlight the important role of the flat spectrum, which tends to emerge observationally even in the presence of non-linear effects which naively may have distorted it.
A similar conclusion was pointed out \citep[][]{keshet2017analytic} based on both non-relativistic and ultra-relativistic shocks in 1D.
It is interesting to mention, in this context, that a flat spectrum naturally arises in 3D if the microphysical plasma configuration is assumed to be self-similar \citep{Katz_etal_07}.

Numerical, in particular ab-initio kinetic simulations of collisionless shocks in 2D play an important role in the theoretical study of the less accessible, 3D problem.
We devote special attention (in \S\ref{sec:2D}) to relativistic shocks in 2D; the results are subsequently generalized for $\Dim\geq3$ (in \S\ref{sec:RelND}).
In particular, an exact relation is derived between the spectral index and the shock-grazing anisotropy parameter $a_1/a_0$, generalizing a 3D result (\KW) to $\Dim=2$ (Eq.~\ref{eq:s_vs_qu}) and to $\Dim\geq3$ (Eq.~\ref{eq:GrazingND}).

In the case of isotropic scattering, the problem of DSA in a relativistic shock can be solved using a rapidly converging expansion in upstream eigenfunctions, as shown with numerically computed eigenfunctions in 3D \citep{Kirk_2000}.
In 2D, the angular eigenfunctions of the transport equation (\ref{eq_transport2}) reduce to the elliptic cosine functions, so the rapidly converging expansion (\ref{eq_DistributionUp}) involves familiar special functions and becomes more transparent.

As in 3D, the first upstream eigenfunction is found to provide an excellent approximation for the angular PDF (Eq.~\ref{eq_FirstEig}), and alone provides an accurate estimate of the spectrum for an arbitrary shock; see Figures \ref{fig:p_index_2D}--\ref{fig:PhaseSpace}.
We show (in \S\ref{sec:RelND}) how the spectrum can be derived directly from this eigenfunction or its approximation (\ref{eq:Phi0ApproxND}).

For isotropic scattering, the spectrum--anisotropy relation can be used to infer an analytic expression for the spectrum, for $\Dim=2$ (Eq.~\ref{eq:s_iso_2D}) and more generally for any $\Dim\geq2$ (Eq.~\ref{eq:GrazingNDiso}), in a method previously invoked in 3D by \KW.
This method relies on two approximations, which appear to be precise or at least very accurate: assuming an analytic behavior of the PDF near the grazing angle, and extrapolating the downstream-frame anisotropy $a_1/a_0$ of non-relativistic shocks (see Eq.~\ref{eq:a1toa0N}) to the relativistic shock regime.
The latter extrapolation is particularly interesting:
while Eq.~(\ref{eq:a1toa0N}) is precise when $\beta_u$ and $\beta_d$ are small, as well as for any $\beta_u$ in the $\beta_d\to0$ limit, it is not a-priori clear that it should remain accurate for large $\beta_d$.

One can show, however, that the deviation $A(\beta_u,\beta_d)$ of $a_1/a_0$ from its extrapolated value is not large.
Consider, for example, ultra-relativistic shocks in 3D, and approximate the PDF using the first upstream eigenfunction, so Eqs.~(\ref{eq:S3D}) and (\ref{eq:Phi0ApproxND}) give $A=(\beta_d/2)(1-\beta_d)/(1+\beta_d)$. While this approximation is inaccurate, it yields $A=0$ in both $\beta_d=0$ and $\beta_d=1$ limits, and a maximal $\sim 11\%$ deviation found at $\beta_d=1/2$.
(Note that while the $\beta_d>1/3$ regime invoked here is not physical, our analysis remains valid for any $0\leq\beta_d<\beta_u<1$.)
The limit where both $\beta_d\to1$ and $\beta_d<\beta_u\to1$ is particularly useful, because in the $\{\beta_u,\beta_d\}$ phase space it is diametrically opposed to the non-relativistic case, and because one may evaluate analytically the overlap between the first eigenfunctions upstream and downstream by invoking the ultra-relativistic approximation in both frames; this yields $(\beta_u-\beta_d)s_p\simeq2$, in agreement with Eq.~(\ref{eq:S3D}), suggesting that the extrapolation is exact.
As another example, notice that the diverging, $s\to\infty$ spectral index expected when $\beta_d\to\beta_u$ so the shock weakens and disappears, implies that $A$ cannot become too negative, $A(\beta,\beta)>-(3/2-\beta^2/2)/(1-\beta^2)$ for any $0\leq\beta<1$.
Moreover, requiring in this limit that the shock frame $\partial_\tau q_s(\mu\simeq0)$ reverses sign at the $\tau=0$ shock front \citep[see {\KW} and Eq.~(4) of][]{Keshet06} yields $0\geq A_s(\beta,\beta)= -(\beta/2)(1-\beta^2)+O(s^{-1})>-0.2$, again vanishing for both $\beta=0$ and $\beta= 1$.
Here, we approximated $a_3\propto -\gamma^2\beta a_2$; \emph{cf.} Eq.~(\ref{eq:diffusion_coeffs}).
In conclusion, the extrapolation is precise, or at least quite accurate, throughout the boundary of the relevant phase space.

The above arguments can be directly generalized for 2D, where the elementary nature of the eigenfunctions renders it easier to test the extrapolation also inside the boundary, as illustrated by Figure \ref{fig:PhaseSpace}.
For example, in the $\beta_d\to 1$ and $\beta_d<\beta_u\to 1$ limit, we obtain $(\beta_u-\beta_d)s_p\simeq 1$, in agreement with Eq.~(\ref{eq:s_iso_2D}), suggesting that the extrapolation here too is exact.
The downstream anisotropy associated with the first upstream eigenfunction can be directly evaluated in 2D for any shock; its deviation from Eq.~(\ref{eq:a1toa0N}) is less than $35\%$ for any $0\leq\beta_d<0.5<\beta_u$.
However, the anisotropy inferred from the first-eigenfunction approximation is inaccurate even at small $\beta_u$.
Better convergence is obtained by considering the spectrum computed in the eigenfunction method, as shown in Figure \ref{fig:PhaseSpace} using the single-eigenfunction overlap (\ref{eq_zero-overlap}), indicating that the extrapolation is accurate to better than a percent throughout the physical regime. The extrapolation is likely to be accurate also at high $\beta_d$, where the eigenfunction method appears to be less adequate.

More important, however, than the above indications in support of the extrapolation, is its success in accounting for the spectrum computed using alternative, numerical or semi-analytic methods, for an arbitrarily relativistic shock and any EOS, as shown for 3D in \KW.
The success of this approach in accounting for the spectrum also in 2D, as we demonstrate in Figure \ref{fig:p_index_2D}, further supports the extrapolation and its validity in both 2D and 3D.
Although the extrapolated downstream anisotropy (\ref{eq:a1toa0N}), in particular $a_1/a_0=\beta_u-\beta_d/2$ in 3D, was not yet justified analytically, we conclude that it may be safely used as a robust diagnostic in shock studies.

Ab-initio particle-in-cell simulations of highly-relativistic shocks in 2D have been able to resolve the onset of particle acceleration, giving rise to nonthermal tails which are consistent with spectral indices in the range $\Se \simeq 2.3\text{--}2.5$ \citep{Spitkovsky2008, sironi2013maximum}, somewhat softer than the $\Se\simeq 2.22$ anticipated in 3D.
For isotropic scattering, in the ultra-relativistic shock limit we find (see Eq.~\ref{eq:sE_Urel_2D}) that $\Se\to (1+\sqrt{13})/2 \simeq 2.303$, possibly accounting for this discrepancy.
Future kinetic simulations in 2D, expected to resolve the spectrum much more accurately, could be compared more carefully to our results.

We find that in 2D, as in 3D, the spectrum in the ultra-relativistic shock limit does not depend on the equation of state.
This differs markedly from the 1D case \citep[see][]{keshet2017analytic}, which thus appears to be the exception.

Our results are useful as a tool for validating numerical simulations in any dimension.
In particular, we present three different methods to infer the spectral index even from an approximate PDF: (i) through the spectrum--grazing anisotropy connection; (ii) by approximating the first downstream eigenfunction; and (ii) through the energy gain and escape probability.
The usefulness of these methods is demonstrated for $\Dim=2$ in \S\ref{sec:otherMethods}, and for $\Dim\geq3$ in \S\ref{sec:RelND}.

\acknowledgments

We thank J. Kirk, A. Spitkovsky, and D. Kagan for helpful discussions.
This research has received funding from the GIF (grant I-1362-303.7/2016), from an IAEC-UPBC joint research foundation grant (grant 300/18), and by the ISF through grant No. 1769/15 and within the ISF-UGC joint research program (grant No. 504/14).

\appendix
\twocolumngrid

\section{Transport equation in 2D}
\label{sec:DiffusionEq}

We consider an infinite, linear (2D version of planar) shock at $z=0$, with flow in the positive $z$ direction.
Relativistic particles with momentum $\bm{p}$ are scattered by electromagnetic modes moving with flow on both sides of the shock. Their PDF $f$ obeys the Fokker--Planck equation, written in the fluid frame as \citep[\eg][]{Blandford_Eichler_87},
\begin{equation}\label{eq_FP}
\frac{\pr f(z,p,\phi)}{\pr t}+\bm{v}\cdot \bm{\nab} f=\frac{\pr}{\pr \bm{p}}\left(\bm{D_{pp}} \frac{\pr f}{\pr  \bm{p}}\right) \fin
\end{equation}
Here, $\bm{{v}}$ is the particle velocity, $\phi$ is the direction of its momentum with respect to the $z$ axis, and $\bm{D_{pp}}$ is the momentum-space diffusion tensor. The diffusion tensor is defined as
\begin{align}
   &\bm{D_{pp}}\equiv
    \begin{pmatrix}
         \myD_{pp}& p \myD_{p\phi}\\
         p \myD_{\phi p} & p^2 \myD_{\phi\phi}
    \end{pmatrix} \,; \\
    &\myD_{kl}\equiv  \left\langle \frac{\Delta {k}\Delta {l}}{2 \Delta t} \right\rangle
    \equiv \frac{1}{2 \Delta t}\int \Delta k \Delta l\, \psi \left(\bm{p},\Delta \bm{p} \right) d(\Delta \bm{p})   \comma\nonumber
\end{align}
where $\psi (\bm{p},\Delta \bm{p}) d(\Delta \bm{p})$ is the element of probability
of a particle changing its momentum from $\bm{p}$ to $\bm{p+\Delta p}$ in time $\Delta t$, and indices $l$ and $k$ represent $p$ or $\phi$.

Assuming that $f(z,p,\phi)$ depends spatially only on the distance from the shock front, and approximating the velocity of the accelerated particles by $c$, the second term on the LHS of Eq.~(\ref{eq_FP}) becomes $c\, \mu\,\pr_{z} f$.
Assuming elastic scattering in the fluid frame, momentum-space diffusion has contributions only from the $\myD_{\phi\phi}$ component.
Equation (\ref{eq_FP}) thus becomes
\begin{align}
    \frac{\pr f}{\pr t}+c\, \mu\frac{\pr f}{\pr z}&=
   \frac{\pr}{\pr \bm{p}}\left[
    \begin{pmatrix}
         0&0\\
         0 & p^2 \myD_{\phi\phi}
    \end{pmatrix}
     \begin{pmatrix}
    \frac{\pr f}{\pr p} \nonumber\\
    \frac{1}{p}\frac{\pr f}{\pr \phi}
    \end{pmatrix}\right]\\&=\frac{\pr }{\pr \phi} \left(\myD_{\phi\phi} \frac{\pr f}{\pr \phi} \right) \fin
\end{align}

Assuming a steady state in the shock frame, and switching to a mixed coordinate system where $z$ is measured in the shock frame,
we obtain
\begin{equation}
    c\,\gamma (\beta +\mu)\frac{\pr f(z_s,p,\phi)}{\pr z_s}=\frac{\pr}{\pr \phi} \left(\myD_{\phi\phi} \frac{\pr f}{\pr \phi} \right) \fin
\end{equation}
where the sub-script $s$ refers to shock-frame variables.

\section{Maxwell-J\"uttner Distribution in 2D}
\label{sec:MJ}
The PDF of an arbitrarily relativistic ideal-gas, known as the Maxwell-J\"{u}ttner distribution, is given (in any dimension) by $f(x^\mu,p^\mu)=Ae^{-(\Theta u_\mu p^\mu)}$ \citep[\eg][]{groot1980relativistic, cercignani2012relativistic}, where $A$ is a temperature-dependent normalization, $u_{\mu}$ is the covariant velocity (three-vector in 2D, henceforth) of the system, $x^\mu$ is the position, $p^{\mu}$ is the momentum, $\Theta\equiv(k_B T)^{-1}$ is the thermodynamic inverse temperature, $k_B$ is the Boltzmann constant, and $T$ is the temperature in Kelvin.
We derive $f$ for the 2D case below.
In Appendix \S \ref{sec:JS_eos} we derive the corresponding EOS, which is found to be simpler than in the 1D or 3D cases, and the resulting jump conditions across a strong shock.

We consider a flat space-time of $\Dim=2$ spatial dimensions, governed by the metric $\eta^{\mu\nu}$ with the sign convention of $\{+,-,-\}$.
The distribution function for an arbitrarily-relativistic gas can be derived by minimizing the free energy of the system for a conserved particle number\citep[\eg][]{hakim2011introduction}.
The free energy density is given by
\begin{equation}
\label{eq_FreeE}
F=\mathcal{\epsilon}-k_BT \sigma \comma
\end{equation}
where $\mathcal{\epsilon}$ is the internal energy density and $\sigma$ is the entropy density.
We may equivalently minimize $F$ for a conserved particle number density $n$,
\begin{equation}\label{eq_delta}
\delta n=0 \comma\quad \delta F=0\fin
\end{equation}

Define
\begin{equation}
\sigma \equiv \frac{u_\mu S^\mu}{c^2} \comma
\end{equation}
\begin{equation}
n\equiv \frac{1}{c^2}\int u_\mu \frac{p^{\mu}}{m} f(p^\mu)  \frac{d^\Dim p}{p_0}mc  \comma \label{eq_dens}
\end{equation}
\begin{equation}
\mathcal{\epsilon} \equiv \frac{1}{c^2} \int \frac{(u_\mu p^\mu)^2}{m} f(p^\mu)  \frac{d^\Dim p}{p_0}mc \comma
\end{equation}
and the entropy-current
\begin{equation}
S^\mu \equiv -\int \frac{p^{\mu}}{m} f(p^\mu) \ln{f(p^\mu)}  \frac{d^\Dim p}{p_0}mc \comma
\end{equation}
where $(d^{\Dim}p/p_0)mc $ is the Lorentz-invariant momentum volume element.
Here, $u_\mu$ is the average
velocity
of the fluid, not to be confused with the individual velocity of each particle (which we denote $p^\mu/m$).
Henceforth, we adopt $\Dim=2$.

Plugging these definition into Eqs.~(\ref{eq_delta}) and (\ref{eq_FreeE}) yields
\begin{equation}
\label{eq_Minimization}
\delta (\Theta F)=\delta \int \left[\Theta(u_\mu p^\mu)^2 + \ln{f(p^\mu)}\right]u_\mu p^\mu f(p^\mu) \frac{d^2p}{c\, p_0}=0 \end{equation}
and
\begin{equation}
\delta n=\delta \int\frac{1}{c} u_\mu p^\mu f(p^\mu) \frac{d^2p}{p_0}=0 \fin
\end{equation}
Introducing a Lagrange multiplier $L$ gives
\begin{equation}
\delta \int\frac{1}{c} \left[\Theta(u_\mu p^\mu) + \ln{f(p^\mu)}+L \right]u_\mu p^\mu f(p^\mu) \frac{d^2p}{p_0}=0 \comma
\end{equation}
from which
\begin{equation}\label{eq_MJ}
f(p^\mu) = A e^{-(\Theta u_\mu p^\mu)} \comma
\end{equation}
where $A$ is a normalization factor.

\section{J\"uttner--Synge equation of state and shock jump conditions in 2D}
\label{sec:JS_eos}

To determine the normalization factor $A$
define the generating function
\begin{equation}\label{eq_GenFun}
\Pi \equiv \int e^{-(\Theta u_\mu p^\mu)}\frac{d^2p}{p_0}mc \comma
\end{equation}
such that
\begin{equation}
\label{eq_nPhi}
n= -\frac{A}{mc^2}\frac{\partial}{\partial \Theta}\Phi \fin
\end{equation}
In the fluid frame $u_0=c$ and $u_{1,2}=0$, so in relativistic polar coordinates
\begin{equation}
    \begin{pmatrix}
    p_0 \\ p_1 \\ p_2
    \end{pmatrix}
    =
   mc \begin{pmatrix}
\cosh\chi \\
\sinh\chi \cos\phi \\
\sinh\chi \sin\phi

    \end{pmatrix}
    \comma
\end{equation}
where $\chi$ is the rapidity and $\phi$ is the angle with respect to the x axis.
Equation (\ref{eq_GenFun}) then becomes
\begin{equation}
\Pi=\int e^{-\Theta m c^2 \cosh\chi} (m c)^2 \sinh\chi \,d\chi \,d\phi=\frac{2\pi m}{\Theta } e^{-\Theta  m c^2} \comma
\end{equation}
so Eq.~(\ref{eq_nPhi}) yields
\begin{equation}
A=\frac{n(\Theta c)^2 }{2\pi(1+\Theta  mc^2)}e^{\Theta  mc^2}\fin
\end{equation}

Next, consider the energy-momentum tensor
\begin{equation}
T^{\mu\nu}\equiv\int \frac{p^\mu p^\nu}{m} f(x,p) \frac{d^2p}{p_0}mc=\frac{A}{m} \frac{\partial^2 \Phi}{\partial(\Theta u_\mu) \partial (\Theta u_\nu)} \fin
\end{equation}
Using the coordinate transformation $\tilde{u}^\mu\equiv\Theta u^\mu$,
\begin{align}
T^{\mu\nu}&=\frac{A}{m}\frac{\partial^2 \Phi}{\partial \tilde{u}_\mu \partial \tilde{u}_\nu}\\
&= \frac{3+3\Theta m c^2+(\Theta mc^2)^2}{1+\Theta mc^2}\frac{n}{\Theta}u^\mu u^\nu -\frac{n}{\Theta}\eta^{\mu\nu}\fin\nonumber
\end{align}
A comparison with
the perfect fluid energy-momentum tensor,
\begin{equation}
T^{\mu\nu}=(\mathcal{\epsilon}+P)u^\mu u^\nu -P\eta^{\mu\nu}\comma
\end{equation}
where $P$ is the pressure, yields the EOS
\begin{gather}
\label{eq_EoS1}
\mathcal{\epsilon}+P=\frac{3+3\Theta m c^2+(\Theta mc^2)^2}{1+\Theta m c^2}\frac{n}{\Theta} \, ;\\
\label{eq_EoS2}
P=\frac{n}{\Theta}\comma
\end{gather}
the latter being the ideal gas law.

The shock jump conditions can now be determined \citep[as in][]{kirk1999particle} from
energy-momentum conservation in the absence of external forces,
\begin{equation}
\label{eq_emCon}
    \nabla T^{\mu\nu}=0\comma
\end{equation}
In the fluid's rest frame, Eq.~(\ref{eq_emCon}) becomes
\begin{equation}
\label{eq_NCon}
\gamma _u \rho _u \beta_u= \gamma _d \rho _d \beta_d \, ;
\end{equation}
\begin{equation}
\label{eq_PCon}
\gamma_u^2 w_u \beta_u^2 + P_u=\gamma_d^2 w_d \beta_d^2 + P_d \, ;
\end{equation}
and
\begin{equation}
\label{eq_ECon}
\gamma_u^2 w_u \beta_u=\gamma_d^2 w_d \beta_d \comma
\end{equation}
where
$\rho\equiv mn$ is the mass density
and
$w\equiv P+\mathcal{\epsilon}$ is the proper enthalpy density.

Assuming a
cold upstream, we neglect the upstream pressure and internal energy, so $w_u/(\rho_u c^2)\simeq 1$, and Eqs.~(\ref{eq_NCon}-\ref{eq_ECon})
give
$\mathcal{\epsilon}_d=\gamma_{r} \rho_d c^2$.
Introducing the adiabatic index $\Ad$ through
\begin{equation}
\label{eq_adiabatic}
P=(\Ad -1)(\mathcal{\epsilon}-\rho c^2) \comma
\end{equation}
then leads to
\begin{equation}\label{eq_enthalpy}
\frac{w_d}{\rho_d c^2}= (\gamma_r-1)\Ad+1
\end{equation}
and the jump condition
\citep[][]{1976PhFl...19.1130B}
\begin{equation}
\label{eq_jump}
\gamma_u^2=\frac{\left(\frac{w_d}{\rho_d c^2}\right)^2(\gamma_r+1)}{ (\gamma_r -1)(2-\Ad)\Ad+2} \fin
\end{equation}

Finally, the above yields
\begin{equation}\label{eq_gamma_zeta}
       \Ad=1+\frac{1+\zeta}{2+\zeta}
\end{equation}
and, assuming a strong shock,
\begin{equation}
    \gamma_r=\frac{\epsilon_d}{\rho_d c^2}=1+\frac{\zeta_d+2}{(1+\zeta_d)\zeta_d} \fin \label{eq_gamma_zeta_2}
\end{equation}
Eqs.~(\ref{eq_gamma_zeta}-\ref{eq_gamma_zeta_2}) can be plugged into Eqs.~(\ref{eq_enthalpy}) and (\ref{eq_jump}), obtaining $\zeta_d$ as a function of $\beta_u$ as the positive root of the polynomial
\begin{align}\label{eq_zeta}
&\beta _u^2\left(2 \zeta _d^3+7 \zeta _d^2+8 \zeta _d+4\right)\left(\zeta _d^2+3 \zeta _d+3\right)^2\nonumber\\&  =\left(3 \zeta _d^3+10 \zeta _d^2+12 \zeta _d+6\right)^2
\end{align}
Equation (\ref{eq_gamma_zeta_2}) then gives
\begin{equation}
      1+\frac{\zeta_d+2}{(1+\zeta_d)\zeta_d}=(1-\beta _d \beta _u) \gamma_d \gamma_u \comma
\end{equation}
which, given $\beta_u$, can be solved for $\beta_d$.

\section{Convergence and Errors}
\label{sec:err}
\begin{figure}[!h]
\centerline{\epsfxsize=8.5cm \epsfbox{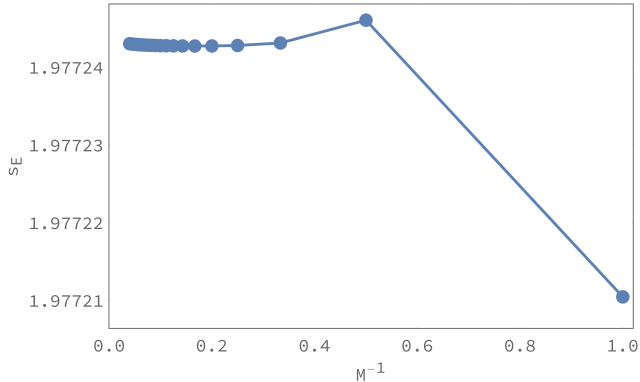}}
\caption{
Convergence plot, showing $\Se$ as a function of $M^{-1}$, with $\gamma_u \beta_u= 1$, calculated with the exact elliptic-cosine functions. Here $M$ is the number of terms the eigenfunction expansion. Each point is extrapolated to infinite resolution in the numerical integration. All points correspond to the JS EOS, and the line is a guide to the eye.
}
\label{fig:Convergence_2D}
\end{figure}

\begin{figure}[!h]
\centerline{\epsfxsize=8.5cm \epsfbox{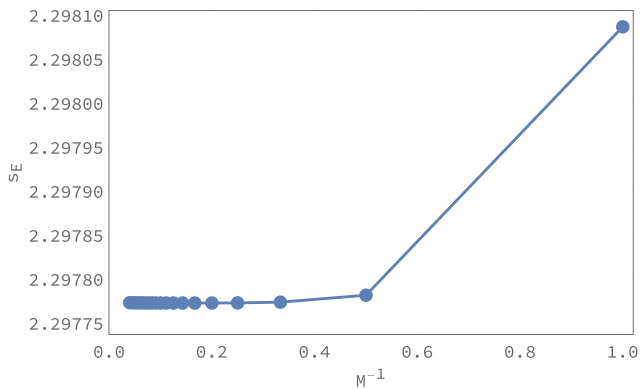}}
\caption{
Same as Figure \ref{fig:Convergence_2D}, but for $\gamma_u \beta_u= 100$, and using the approximation
(\ref{eq_ceApprox}).
}
\label{fig:Convergence_Approx_2D}
\end{figure}

We demonstrate the convergence of the expansion in upstream eigenfunctions
by
varying the number $M$ of terms in the expansion and in the numerical integration resolution.
The spectral index, extrapolated to infinite resolution,
is shown in Figures \ref{fig:Convergence_2D} and \ref{fig:Convergence_Approx_2D}
as a function of $M^{-1}$, for
the exact elliptic-cosine functions with
$\gamma_u\beta_u=1$ and for the ultra-relativistic shock approximation (\ref{eq_ceApprox}) with $\gamma_u\beta_u=100$, respectively.
The errors bars presented in Figure \ref{fig:p_index_2D} are estimated
by extrapolating the data
to $M\to\infty$.

\bibliography{DSA}
\bibliographystyle{apj}

\end{document}